\documentstyle[aps,12pt]{revtex}

\begin{document}

\preprint{EFUAZ FT-97-50-REV}

\title{Additional Equations Derived from the Ryder Postulates
in the $(1/2,0)\oplus (0,1/2)$ Representation of
the Lorentz Group\thanks{Submitted to ``Nuovo Cimento A"}}

\author{{\bf Valeri V. Dvoeglazov}}

\address{Escuela de F\'{\i}sica, Universidad Aut\'onoma de Zacatecas\\
Apartado Postal C-580, Zacatecas 98068 Zac., M\'exico\\
Internet address: valeri@cantera.reduaz.mx\\
URL: http://cantera.reduaz.mx/\~~valeri/valeri.htm
}

\date{October 27, 1997; Revised: September 1998}

\maketitle

\bigskip

\begin{abstract}
Developing recently proposed constructions for
the description of  particles in the $(1/2,0)\oplus (0,1/2)$
representation space, we derive the second-order equations.
The similar ones were proposed in the sixties and the seventies
in order to understand the nature of various mass and spin states
in the representations of the $O(4,2)$ group. We give some
additional insights into this problem.  The used procedure can be
generalized for {\it arbitrary} number of lepton families.
\end{abstract}

\bigskip


A correct equation for  an adequate description of  neutrinos
was sought for a long time~\cite{LY,Tok,SG,Fush1,Simon}. This problem
is, in general, connected with the problem of taking the massless limit of
relativistic equations. For instance, it has been known for a long time
that {\it ``one cannot simply set the mass equal to zero in a manifestly
covariant massive-particle equation, in order to obtain the corresponding
massless case"}, e.~g., ref.~[5a].

Secondly, in the seventies the second-order equation in the
4-dimensional representation of the $O(4,2)$ group was proposed by Barut
{\it et al.} in order to solve the problem of the mass splitting of
leptons~\cite{Bar0,Wilson,Bar} and by Fushchich {\it et al.}, for
describing various spin states in this representation~\cite{Fush,Fush2}.
The equations (they proposed) may depend on two parameters. Recently we
derived the Barut-Wilson equation on the basis of the first
principles~\cite{DVB}.\footnote{Briefly, the scheme for derivation of the
equation
\begin{equation}
\left [i\gamma^\mu \partial_\mu  + \alpha_2 \partial^\mu
\partial_\mu  -\kappa \right ] \phi (x^\mu)
= 0\,\,   \label{Barut}
\end{equation}
is the following. First, apply the generalized Ryder-Burgard relation
(see below, Eq. (\ref{RB})) and the standard scheme for
the derivation  of relativistic wave equations~\cite[footnote \# 1]{AV}.
Then form the  Dirac 4-spinors; the left- and right parts of them are
connected as follows:
\begin{mathletters} \begin{eqnarray}
\phi_{_L}^\uparrow (p^\mu) &=& - \Theta_{[1/2]} [\phi_{_R}^{\downarrow}
(p^\mu)]^\ast \quad,\quad \phi_{_L}^\downarrow (p^\mu) = + \Theta_{[1/2]}
[\phi_{_R}^{\uparrow}(p^\mu)]^\ast \,\, ,\label{1a}\\
\phi_{_R}^\uparrow (p^\mu) &=& -
\Theta_{[1/2]} [\phi_{_L}^{\downarrow} (p^\mu)]^\ast \quad,\quad
\phi_{_R}^\downarrow (p^\mu) = + \Theta_{[1/2]}
[\phi_{_L}^{\uparrow} (p^\mu)]^\ast \label{1b}\,\,,
\end{eqnarray}
\end{mathletters}
in order to obtain
\begin{equation}
\left [a \,{i\gamma^\mu \partial_\mu \over m}
+b\, {\cal C} {\cal K} - \openone\right ] \Psi (x^\mu) = 0\,\, ,\label{de}
\end{equation}
in the coordinate space.
Transfer to the Majorana
representation with the unitary matrix
\begin{equation} U ={1\over
2}\pmatrix{\openone -i\Theta_{[1/2]} & \openone +i\Theta_{[1/2]}\cr
-\openone -i\Theta_{[1/2]} & \openone -i\Theta_{[1/2]}\cr}\quad,\quad
U^\dagger = {1\over 2}\pmatrix{\openone -i\Theta_{[1/2]} & -\openone
-i\Theta_{[1/2]}\cr \openone + i\Theta_{[1/2]} & \openone
-i\Theta_{[1/2]}\cr}\,\,.\label{maj}
\end{equation}
Finally, one obtains
the set
\begin{mathletters}
\begin{eqnarray}
\left [ a {i\gamma^\mu \partial_\mu \over m} -\openone \right ] \phi - b
\,\chi &=& 0 \,\,,\\
\left [ a {i\gamma^\mu \partial_\mu \over m} -\openone \right ]\chi  - b
\,\phi &=& 0\,\,
\end{eqnarray}
\end{mathletters}
for $\phi (x^\mu) = \Psi_1 +\Psi_2$ or $\chi (x^\mu)=\Psi_1 -\Psi_2$
(where $\Psi^{^{MR}} (x^\mu) = \Psi_1 +i\Psi_2$).
With the identification $a/2m \rightarrow \alpha_2$ and $m(1-b^2)/2a
\rightarrow \kappa$ the above set leads to the second-order equation of
the Barut type.}

Thirdly, we found the possibility of generalizations of the $(1,0)\oplus
(0,1)$ equations (namely, the Maxwell's equations and the
Weinberg-Tucker-Hammer equations\footnote{In general, the latter does
{\it not} completely reduce to the former after taking the massless limit
in the ``accustomed" way.}) also on the basis of including two independent
constants~\cite{Dvo}, cf.  also~\cite{Dva}.  This induced us to look
inside the problem on the basis of the first principles; my research was
started in~\cite{Dvo1}.

In this paper, we first apply the Ahluwalia
reformulation~\cite{AV,DVON,DVON1} of the Majorana-McLennan-Case construct
for neutrino~\cite{MAJOR,MLC} with the purpose of the derivation of
relevant equations,  we recalled above.

The following definitions and postulates are used:
\begin{itemize}

\item
The operators of the discrete symmetries are defined as follows:
a) the space inversion operator:
\begin{eqnarray}
S^s_{[1/2]} = \pmatrix{0&\openone\cr
\openone & 0\cr}\, ,
\end{eqnarray}
is the $4\times 4$ anti-diagonal matrix;
b) the charge conjugation operator:
\begin{eqnarray}
S^c_{[1/2]} = \pmatrix{0 & i\Theta_{[1/2]}\cr
-i\Theta_{[1/2]} & 0\cr}{\cal K}\,\, , \label{cco}
\end{eqnarray}
with ${\cal K}$ being the operation of complex conjugation; and
$\left (\Theta_{[j]}\right )_{h,\,h^\prime} = (-1)^{j+h}
\delta_{h^\prime,\,-h}$  being the Wigner operator.

\item
The left- ($\phi_{_L}$ and $\zeta \Theta_{[j]}\phi_{_R}^\ast$)
and the right-  ($\phi_{_R}$ and $\zeta^\prime
\Theta_{[j]}\phi_{_L}^\ast$) spinors  are transformed to the frame with
the momentum $p^\mu$ (from the zero-momentum frame) as follows:
\begin{mathletters}
\begin{eqnarray}
\phi_{_R} (p^\mu)\, &=& \,\Lambda_{_R} (p^\mu \leftarrow
\overcirc{p}^\mu)\,\phi_{_R} (\overcirc{p}^\mu) \, = \, \exp (+\,{\bf J}
\cdot {\bbox \varphi}) \,\phi_{_R} (\overcirc{p}^\mu)\,\,,\\
\phi_{_L}  (p^\mu)\, &=&\, \Lambda_{_L} (p^\mu \leftarrow
\overcirc{p}^\mu)\,\phi_{_L} (\overcirc{p}^\mu) \, = \, \exp (-\,{\bf J}
\cdot {\bbox \varphi})\,\phi_{_L} (\overcirc{p}^\mu)\,\,.\label{boost0}
\end{eqnarray}
\end{mathletters}
$\Lambda_{_{R,L}}$ are the matrices for the Lorentz boosts; ${\bf J}$ are
the spin matrices for spin $j$, e.~g., ref.\cite{Var}; ${\bbox \varphi}$
are parameters of the given boost.  If we restrict ourselves by the case of
bradyons they are defined, {\it e.~g.}, refs.~\cite{Ryder,Dva}, by means
of:
\begin{equation}\label{boost} \cosh (\varphi) =\gamma =
\frac{1}{\sqrt{1-v^2}} = \frac{E}{m},\quad \sinh (\varphi) = v\gamma =
\frac{\vert {\bf p}\vert}{m},\quad \hat {\bbox \varphi} = {\bf n} =
\frac{{\bf p}}{\vert {\bf p}\vert}\,\, .
\end{equation}

\item
The Ryder-Burgard relation between spinors in the zero-momentum
frame\footnote{This name was introduced by D. V. Ahluwalia  when
considering the $(1,0)\oplus (0,1)$ representation~\cite{Dva}. If one uses
$\phi_{_R} (\overcirc{p}^\mu) = \pm \phi_{_L} (\overcirc{p}^\mu)$, cf.
also~\cite{Faust,Ryder}, after application of the Wigner rules for the
boosts of the 3-component objects to the momentum $p^\mu$, one immediately
arrives at the ``Bargmann-Wightman-Wigner type quantum field theory",
ref.~\cite{BWW} (cf.  also the old papers~\cite{Nig,Gel,Sokol} and the
recent papers~\cite{ZIINO,DVOED}), in this representation.
The reader can still reveal some terminological obscurities
in~\cite{Dva}.} is established
\begin{equation} \phi_{_L}^h
(\overcirc{p}^\mu) = a (-1)^{{1\over 2} - h} e^{i(\vartheta_1
+\vartheta_2)} \Theta_{[1/2]} [\phi_{_L}^{-h} (\overcirc{p}^\mu)]^\ast + b
e^{2i\vartheta_h} \Xi^{-1}_{[1/2]} [\phi_{_L}^h (\overcirc{p}^\mu)]^\ast
\,\, ,\label{RB}
\end{equation}
with the {\it real} constant $a$ and $b$ being arbitrary at this stage.
$h$ is a quantum number corresponding to the helicity,
\begin{eqnarray}
\Xi_{[1/2]} = \pmatrix{e^{i\phi}&0\cr
0&e^{-i\phi}\cr}\,\, ,
\end{eqnarray}
$\phi$ is here the azimuthal angle related to ${\bf p} \rightarrow {\bf
0}$; in general, see the cited papers for  the notation.\footnote{In
general, one can connect also $\phi_L^\uparrow$ and $\phi_L^\downarrow$.
with using the $\Omega$ matrix (see formulas (22a,b) in ref.~\cite{AV}):
\begin{equation}
\phi_L^\uparrow (\overcirc{p}^\mu) = \Omega \phi_L^\downarrow
(\overcirc{p}^\mu)\,\, , \quad
\Omega =\pmatrix{ cotan (\theta/2) & 0\cr
0& - tan (\theta/2)\cr } = {\vert {\bf p}\vert \over \sqrt{ {\bf p}^{\,2}
- p_3^2} } (\sigma_3 + {p_3 \over \vert {\bf p} \vert })\,\, .
\end{equation}
We did not yet find the explicitly covariant form of the resulting
equation.}

\item
One can form either Dirac 4-spinors:
\begin{equation}
u_h (p^\mu) =\pmatrix{\phi_{_R} (p^\mu)\cr
\phi_{_L} (p^\mu)\cr}\quad,\quad
v_h (p^\mu) = \gamma^5 u_h (p^\mu)\,\, ,
\end{equation}
or the second-type spinors~\cite{AV}, see also~\cite{Sokol,DVOED}:
\begin{equation}
\lambda (p^\mu) = \pmatrix{(\zeta_\lambda \Theta_{[j]}) \phi_{_L}^\ast
(p^\mu) \cr \phi_{_L} (p^\mu)\cr}\quad,\quad
\rho (p^\mu) = \pmatrix{\phi_{_R}
(p^\mu) \cr (\zeta_\rho \Theta_{[j]})^\ast \phi_{_R}^\ast
(p^\mu)\cr}\,\, ,
\end{equation}
or even more general forms of 4-spinors
depending on the phase factors between their left- and right- parts and
helicity sub-spaces that they belong to.
For the second-type spinors, the author of ref.~\cite{AV} proposed
several forms of the field operators, e.~g.,
\begin{eqnarray}
\nu^{DL} (x^\mu) &=& \sum_\eta \int \frac{d^3 {\bf p}}{(2\pi)^3}
{1\over 2E_p} \left [ \lambda^S_\eta (p^\mu) c_\eta (p^\mu) \exp (-ip\cdot
x) +\right.\nonumber\\
&+&\left.\lambda^A_\eta (p^\mu) d_\eta^\dagger (p^\mu) \exp
(+ip\cdot x)\right ]\,\, .
\end{eqnarray}

\end{itemize}

On the basis of these definitions on using the standard
rules~\cite[footnote \# 1]{AV}  one can  derive:

\begin{itemize}

\item
In the case $\vartheta_1 =0$, $\vartheta_2 =\pi$ the following
equations are obtained for $\phi_{_L} (p^\mu)$ and
$\chi_{_R} = \zeta_\lambda \Theta_{[1/2]} \phi_{_L}^\ast (p^\mu)$
:\footnote{The phase factors $\zeta$ are defined by various constraints
imposed on the 4-spinors (or corresponding operators), e.~g.,
the condition of the self/anti-self charge conjugacy gives
$\zeta_\lambda^{S,A}=\pm i$. But, one should still note that
phase factors also depend on the phase factor in the definition
of the charge conjugation operator (\ref{cco}). The ``mass term" of
resulting dynamical equations may also be different.}
\begin{mathletters}
\begin{eqnarray}
\phi_{_L}^h (p^\mu ) &=& \Lambda_{_L} (p^\mu \leftarrow \overcirc{p}^\mu)
\phi_{_L}^h (\overcirc{p}^\mu)
= {a \over \zeta_\lambda}
(-1)^{{1\over 2} +h} \Lambda_{_L} (p^\mu \leftarrow \overcirc{p}^\mu)
\Lambda_{_R}^{-1} (p^\mu \leftarrow \overcirc{p}^\mu) \chi_{_R}^h (p^\mu)+
\nonumber\\
&+&{b\over \zeta_\lambda} \Lambda_{_L} (p^\mu \leftarrow \overcirc{p}^\mu)
\Xi^{-1}_{[1/2]} \Theta^{-1}_{[1/2]} \Lambda_{_R}^{-1} (p^\mu \leftarrow
\overcirc{p}^\mu) \chi_{_R}^{-h} (p^\mu)\,\,,\\
\chi_{_R}^{-h} (p^\mu) &=& \Lambda_{_R}
(p^\mu \leftarrow \overcirc{p}^\mu) \chi_{_R}^{-h}
(\overcirc{p}^\mu) =
a \zeta_\lambda (-1)^{{1\over 2}
-h} \Lambda_{_R} (p^\mu \leftarrow \overcirc{p}^\mu) \Lambda_{_L}^{-1}
(p^\mu \leftarrow \overcirc{p}^\mu) \phi_{_L}^{-h} (p^\mu) +\nonumber\\
&+& b\zeta_\lambda \Lambda_{_R} (p^\mu \leftarrow \overcirc{p}^\mu)
\Theta_{[1/2]} \Xi_{[1/2]} \Lambda_{_L}^{-1} (p^\mu \leftarrow
\overcirc{p}^\mu) \phi_{_L}^{h} (p^\mu)\,\,.
\end{eqnarray}
\end{mathletters}
Hence, the equations for the 4-spinors
$\lambda^{S,A}_\eta (p^\mu)$
take the forms:\footnote{$\eta$ is the chiral helicity
quantum number introduced in ref.~\cite{AV}.}
\begin{mathletters}
\begin{eqnarray}
ia
{\widehat p \over m} \lambda^S_\uparrow (p^\mu) - (b{\cal C}{\cal K}
-\openone) \lambda^S_\downarrow (p^\mu) &=& 0\,\, ,
\label{m1}\\
ia {\widehat p \over m} \lambda^S_\downarrow (p^\mu) + (b
{\cal C}{\cal K} -\openone) \lambda^S_\uparrow (p^\mu) &=& 0\,\, ,
\label{m2}\\
ia {\widehat p \over m} \lambda^A_\uparrow (p^\mu) - (b
{\cal C}{\cal K} +\openone) \lambda^A_\downarrow (p^\mu) &=& 0\,\, ,
\label{m3}\\
ia {\widehat p \over m} \lambda^A_\downarrow (p^\mu) + (b {\cal
C}{\cal K} +\openone) \lambda^A_\uparrow (p^\mu) &=& 0\,\,
\label{m4},
\end{eqnarray} \end{mathletters}
$a =\pm (b-1)$ if we want to have $p_0^2 - {\bf p}^2 = m^2$ for massive
particles.

\item
We can write several forms of equations in the coordinate representation
depending on the relations between creation/annihilation operators.
For instance,
provided that we imply $d_\uparrow (p^\mu) = +ic_\downarrow (p^\mu)$
and $d_\downarrow (p^\mu) = -ic_\uparrow (p^\mu)$; the ${\cal K}$
operator acts on $q-$ numbers as hermitian conjugation,
the first generalized equation in the coordinate space reads
\begin{equation}
\left [ ia {\gamma^\mu \partial_\mu \over m} - (b-\openone) \gamma^5 {\cal
C} {\cal K} \right ] \Psi (x^\mu) = 0\,\, .
\end{equation}

Transferring into the Majorana representation one obtains two
real equations:\footnote{It seems that this procedure can be carried out
for any spin, cf.~\cite{Dvo2}.}
\begin{mathletters} \begin{eqnarray}
ia {\gamma^\mu \partial_\mu \over m} \Psi_1 (x^\mu) -i (b-\openone)
\gamma^5 \Psi_2 (x^\mu) &=& 0\,\, ,\\
ia {\gamma^\mu \partial_\mu \over
m} \Psi_2 (x^\mu) -i (b-\openone)\gamma^5 \Psi_1 (x^\mu) &=& 0\,\, .
\end{eqnarray} \end{mathletters}
for real and imaginary parts of the field
function $\Psi^{^{MR}} (x^\mu) = \Psi_1 (x^\mu) +i\Psi_2 (x^\mu)$.
In the case of $a =1-b$
and considering the field function $\phi= \Psi_1 +\Psi_2$
we come to the Sokolik equation
for the spinors of the {\it second
kind}~\cite[Eq.(8)]{Sokol} and ref.~\cite[Eqs.(14,18)]{DVOED}.
Next, we  come to the  second-order equation in the coordinate
representation for massive particles
\begin{equation} \left [ a^2
{\partial_\mu \partial^\mu \over m^2} +(b-1)^2 \right ] \cases{\Psi_1
(x^\mu) &\cr \Psi_2 (x^\mu) &\cr} = 0\,\,.\label{kg} \end{equation} Of
course, it may be reduced to the Klein-Gordon equation.
In general, there may exist mass splitting between various $CP-$
conjugate states. We shall return to this question in other papers.

\item
One can find the relation between creation/annihilation
operators for another equation ($\beta_1 , \, \beta_2 \in \Re e$)
\begin{equation}
\left [ ia {\gamma^\mu \partial_\mu \over m} - e^{i\alpha_1}\beta_1
\gamma^5 {\cal C} {\cal K}  + e^{i\alpha_2} \beta_2 \right ] \Psi (x^\mu)
= 0\,\, ,\label{neq1}
\end{equation}
which would be consistent with the equations
(\ref{m1}-\ref{m4}).\footnote{As one can expect from  this consideration
the equation (\ref{neq1}) may be reminiscent of the works of the 60s,
refs.~\cite{SG,Fush1,Simon,Rasp}.}
Here they are:
\begin{mathletters}
\begin{eqnarray}
(b-1) c_\uparrow &=& ie^{i\alpha_1}\beta_1 d_\downarrow -ie^{i\alpha_2}
\beta_2 c_\downarrow\,\, ,\\
(b-1) c_\downarrow &=& -ie^{i\alpha_1}\beta_1 d_\uparrow +ie^{i\alpha_2}
\beta_2 c_\uparrow\,\, ,\\
(b-1) d_\uparrow^\dagger &=& -ie^{i\alpha_1}\beta_1 c_\downarrow^\dagger
-ie^{i\alpha_2} \beta_2 d_\downarrow^\dagger\,\, ,\\
(b-1) d_\downarrow^\dagger &=& ie^{i\alpha_1}\beta_1 c_\uparrow^\dagger
+ie^{i\alpha_2} \beta_2 d_\uparrow^\dagger\,\, .
\end{eqnarray}
\end{mathletters}
The condition of the compatibility ensures that $\alpha_2 =0,\pi$ and
$\beta_1^2 +\beta_2^2 = (b-1)^2$. We assumed that two
annihilation operators are linear independent. If $\beta_1 =0$, we
recover the Dirac equation but with additional constraints put on the
creation/annihilation operators, $c_\uparrow = \mp ic_\downarrow$ and
$d_\uparrow = \pm i d_\downarrow$. The phase phactor $\alpha_1$ remains
unfixed at this stage.

In the Majorana representation the resulting set of the real equations
are
\begin{mathletters}
\begin{eqnarray}
&&\left [ia {\gamma^\mu \partial_\mu \over m} +i\beta_1 \sin\alpha_1
\gamma^5  +\beta_2 \right ] \Psi_1 -i\beta_1 \cos\alpha_1 \gamma^5 \Psi_2
= 0\,\, ,\\
&&\left [ia {\gamma^\mu \partial_\mu \over m} -i\beta_1 \sin\alpha_1
\gamma^5  +\beta_2 \right ] \Psi_2 -i\beta_1 \cos\alpha_1 \gamma^5 \Psi_1
= 0\,\, .
\end{eqnarray} \end{mathletters}
For instance in the $\alpha_1 = {\pi \over 2}$ we obtain
\begin{mathletters}
\begin{eqnarray}
\left [ ia {\gamma^\mu \partial_\mu \over m} +i \beta_1 \gamma^5 +\beta_2
\right ] \Psi_1 &=& 0\,\, ,\label{gf1}\\
\left [ ia {\gamma^\mu \partial_\mu \over m} -i \beta_1 \gamma^5 +\beta_2
\right ] \Psi_2 &=& 0\,\, \label{gf2}.
\end{eqnarray}
\end{mathletters}
But, in any case one can recover the
Klein-Gordon equation for both real and imaginary parts of the field
function, Eq. (\ref{kg}).  It is not yet clear, if the constructs
discussed recently in ref.~\cite{Rasp} are permitted.

\item
But, we are able to consider other constraints on the
creation/annihilation operators, introduce various types of
fields operators (as in~\cite{Dvo}) and/or generalize even more the
Ryder-Burgard relation (see footnote \# 4 of the present paper,
for instance). In the general case, we suggest to start from
\begin{equation}
(a{\widehat p \over m} -\openone) u_h (p^\mu) +i b (-1)^{{1\over 2}-h}
\gamma^5 {\cal C} u_{-h}^\ast (p^\mu) =0\,\,;
\end{equation}
i.~e., the equation (11) of~\cite{DVB}. But, as opposed to the cited paper,
we write the coordinate-space equation in the form:
\begin{equation}
\left [a \,{i\gamma^\mu \partial_\mu \over m}
+b_1\, {\cal C} {\cal K} - \openone\right ] \Psi (x^\mu)
+ b_2 \gamma^5  {\cal C} {\cal K} \widetilde{\Psi} (x^\mu)
= 0\,\, ,\label{dem}
\end{equation}
thus introducing the third parameter. Then we can perform
the same procedure as in ref.~\cite{DVB}. Implying $\Psi^{^{MR}}
=\Psi_1 + i\Psi_2$ and $\widetilde \Psi^{^{MR}}
=\Psi_3 + i \Psi_4$, one obtains real equations in the Majorana
representation
\begin{mathletters}
\begin{eqnarray}
(a {i\gamma^\mu \partial_\mu \over m} - \openone )\phi
-b_1 \chi +ib_2 \gamma^5 \widetilde \phi &=& 0\, ,\\
(a {i\gamma^\mu \partial_\mu \over m} - \openone )\chi
-b_1 \phi -ib_2 \gamma^5 \widetilde \chi &=& 0\, ,
\end{eqnarray}
\end{mathletters}
for $\phi = \Psi_1 +\Psi_2$, $\chi =\Psi_1 - \Psi_2$
and
$\widetilde\phi = \Psi_3 +\Psi_4$, $\widetilde\chi =\Psi_3 - \Psi_4$.
After algebraic transformations we have:
\begin{mathletters}
\begin{eqnarray}
&& (a{i\gamma^\mu \partial_\mu \over m} - b_1 -\openone ) \left [ 2ia
{\gamma^\nu \partial_\nu \over m} + a^2 {\partial^\nu \partial_\nu \over
m^2} + b_1^2 - \openone \right ] \Psi_1  -\nonumber\\
&& \qquad - ib_2 \gamma^5 \left [ 2ia
{\gamma^\mu \partial_\nu \over m} - a^2 {\partial^\mu \partial_\mu \over
m^2} - b_1^2 + \openone \right ] \Psi_4 = 0\, ;\\
&& (a{i\gamma^\mu \partial_\mu \over m} + b_1 -\openone ) \left [ 2ia
{\gamma^\nu \partial_\nu \over m} + a^2 {\partial^\nu \partial_\nu \over
m^2} + b_1^2 - \openone \right ] \Psi_2  -\nonumber\\
&& \qquad - ib_2 \gamma^5 \left [ 2ia
{\gamma^\mu \partial_\mu \over m} - a^2 {\partial^\mu \partial_\mu \over
m^2} - b_1^2 + \openone \right ] \Psi_3 = 0\, ,
\end{eqnarray}
\end{mathletters}
the third-order equations. However, the field operator
$\widetilde \Psi$ may be linear dependent on the states
included in the $\Psi$. So, relations may exist
between $\Psi_{3,4}$ and $\Psi_{1,2}$. If we apply
the most simple (and accustomed) constraints
$\Psi_1 = -i\gamma^5 \Psi_4$ and $\Psi_2 =i\gamma^5 \Psi_3$
one should recover the Dirac-Barut-like equation
with {\it three mass eigenvalues}:
\begin{equation}
\left [ i\gamma^\mu \partial_\mu - m {1\pm b_1 \pm b_2 \over a}
\right]\times
\left [ i\gamma^\nu \partial_\nu + {a\over 2m} \partial^\nu \partial_\nu
+ m {b_1^2 -1 \over 2a}\right ] \Psi_{1,2} =0\, .
\end{equation}

Thus, we can conclude that as in several previous works
we observe that the physical results depend on
the stage where one  applies the relevant constraints.
Furthermore, we apparently note that the similar results
can be obtained by consecutive applications of the generalized
Ryder-Burgard relations.

As indicated by Barut himself, several ways for introdcution of
interaction  with 4-vector potential exist in second-order equations.
Only considering the correct one (and, probably, taking into account
$\gamma^5$ axial currents introduced in~\cite{DVON1}), we shall be able to
answer the question of  why the $\alpha_2$ parameter of the Barut works is
fixed by means of the use of the {\it classical} value of anomalous 
magentic moments; and on what physical basis we have to fix other 
parameters we introduced above.

\end{itemize}

In conclusion, we presented a very natural way of deriving the
massive/massless equations in the $(1/2,0)\oplus (0,1/2)$ representation
space, which leads to those given by other researchers in the past.
It is known that present-day neutrino physics has come across
serious difficulties. No experiment and observation are in
agreement with theoretical predictions of the standard model.
Furthermore, Barut's way of solving the hierarchy problem
was almost forgotten, in spite of its simplicity and beauty.
In fact, an idea of imposing certain relations between
convective and rotational motions of a fermion is much more
physical than all other modern-fashioned models.
It may be very fruitful, because as we have shown previously and here,
the $(j,0)\oplus (0,j)$ representation spaces have a very rich
internal structure. I hope the question of whether proposed equations have
some relevance to the description of the physical world, will be solved in
the near future.

{\it Acknowledgments.} The work was motivated by the papers
of Prof. D. V. Ahluwalia, by very useful frank discussions
and phone  conversations with Prof. A. F. Pashkov during last 15 years,
by Prof. A. Raspini who kindly sent me his papers, and by the critical
report of anonymous referee of the  ``Foundation of Physics" which,
nevertheless, was of crucial importance in initiating my present
research. I also acknowledge the discussion with Profs. A. E. Chubykalo,
Y.  S.  Kim, R.  M.  Santilli and Yu. F. Smirnov.
After completing the preliminary version I had fruitful discussions
with Dr. G. Quznetsov. Many thanks to all them.

Zacatecas University, M\'exico, is thanked for awarding
a professorship.  This work has been partly supported by
the Mexican Sistema Nacional de Investigadores and the Programa de Apoyo a
la Carrera Docente.

\end{document}